\documentclass[aps,twocolumn,superscriptaddress,nobibnotes]{revtex4-2}

\usepackage{mhchem}

\usepackage{graphicx}
\usepackage{caption}

\usepackage{siunitx}
\DeclareSIUnit\angstrom{\text {Å}}

\usepackage{hyperref}
\hypersetup{colorlinks=true, citecolor=blue, urlcolor=blue, linkcolor=blue}

\begin{document}
\title{Extra electron reflections in concentrated alloys do not necessitate short-range order}

\author{Flynn Walsh}
\affiliation{Materials Sciences Division, Lawrence Berkeley National Laboratory, Berkeley, CA, USA}
\affiliation{Graduate Group in Applied Science \& Technology, University of California, Berkeley, CA, USA}
\author{Mingwei Zhang}
\affiliation{Materials Sciences Division, Lawrence Berkeley National Laboratory, Berkeley, CA, USA}
\affiliation{National Center for Electron Microscopy, Lawrence Berkeley National Laboratory, Berkeley, CA, USA}
\affiliation{Department of Materials Science \& Engineering, University of California, Berkeley, CA, USA}
\author{Robert O. Ritchie}
\affiliation{Materials Sciences Division, Lawrence Berkeley National Laboratory, Berkeley, CA, USA}
\affiliation{Department of Materials Science \& Engineering, University of California, Berkeley, CA, USA}
\author{Andrew M. Minor}
\affiliation{Materials Sciences Division, Lawrence Berkeley National Laboratory, Berkeley, CA, USA}
\affiliation{National Center for Electron Microscopy, Lawrence Berkeley National Laboratory, Berkeley, CA, USA}
\affiliation{Department of Materials Science \& Engineering, University of California, Berkeley, CA, USA}
\author{Mark Asta}
\affiliation{Materials Sciences Division, Lawrence Berkeley National Laboratory, Berkeley, CA, USA}
\affiliation{Department of Materials Science \& Engineering, University of California, Berkeley, CA, USA}
\affiliation{\rm{Corresponding author: \href{mailto:mdasta@berkeley.edu}{mdasta@berkeley.edu}}}

\begin{abstract}

In many concentrated alloys of current interest, the observation of diffuse superlattice intensities by transmission electron microscopy has been attributed to chemical short-range order.
We briefly review these findings and comment on the plausibility of widespread interpretations, noting the absence of expected peaks, conflicts with theoretical predictions, and the possibility of alternative explanations.

\end{abstract}

\maketitle

The nature of chemical short-range order (SRO) in face-centered cubic (fcc) alloys containing several 3\textit{d} principal elements, such as ``medium-entropy" VCoNi and CrCoNi, has been intensively investigated in recent years \cite{ding18,schonfeld19,kostiuchenko20,zhang20,zhou21,inoue21,walsh21,chen21,chen22,zhou22,yu22,du22,zhang22,hsiao22,su22,ghosh22,li23,zhu23}.
While there may be little evidence that SRO can be controlled to tailor the bulk mechanical properties of these materials \cite{zhou21,inoue21,zhang22,li23}, it has been argued that an essentially ubiquitous degree of \AA-scale order could nonetheless play an important role in a wide range of properties \cite{ding18,walsh21}.
For example, many alloys containing Co are predicted to form hexagonal close-packed lattices at ambient conditions \cite{niu18,dong18}, but quenched-in SRO could account for the persistent metastability of the fcc phase \cite{ding18,walsh21,yu22}.

\begin{figure}
\includegraphics{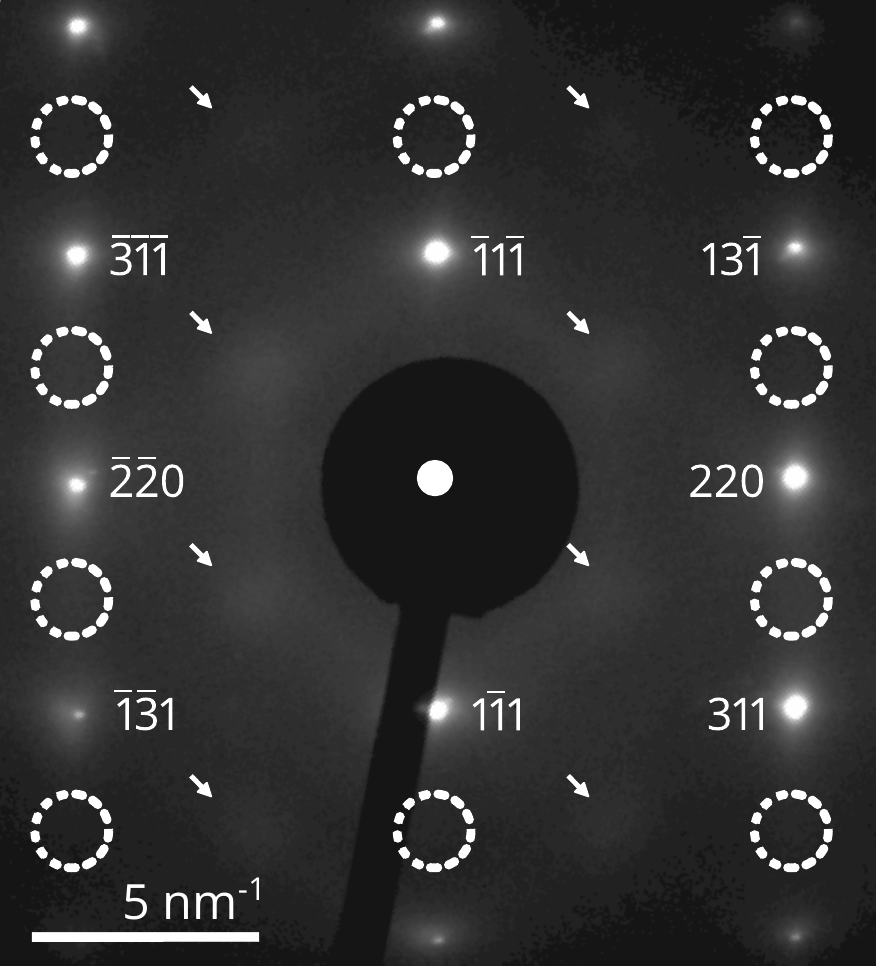}
\caption{\label{fig:diffraction}
    \textbf{Electron diffraction of CrCoNi in the $[\bar{1}12]$ ZA.}
    This pattern, which is based on experimental data from a previous study \cite{zhang22}, is representative of literature results for alloys discussed in the text.
    Diffuse intensities at $\frac{1}{2}\{311\}$ superlattice positions are marked with arrows, but there are no peaks at $\frac{1}{2}\{111\}$ sites, as highlighted by the dotted circles.
}
\end{figure}

This view appears to be corroborated by recent transmission electron microscopy (TEM) purporting the presence of local order in a variety of samples subject to minimal thermal processing beyond high-temperature homogenization.
For example, the observation of SRO in VCoNi was proposed \cite{chen21} on the basis of diffuse intensities at $\frac{1}{2}\{311\}$ superlattice sites in reciprocal space while imaging in the $[\bar{1}12]$ zone axis (ZA), which indicates the crystallographic direction of electron incidence.
An equivalent electron diffraction pattern is shown in Fig. \ref{fig:diffraction}.
Additional $\frac{1}{3}\{422\}$ intensities were later reported in the $[\bar{1}11]$ ZA \cite{chen22}.
Similar observations have been at various points attribtued to SRO in a Cr-Ni-based alloy \cite{kim15}, Mn-Fe-based alloys \cite{seol20,liu21,kayani22,seol22}, CrCoNi \cite{zhou22,zhang22}, CrMnFeCoNi \cite{su22}, and VFeCoNi \cite{zhu23}.
The same features have also been reported without the assumption of SRO \cite{xu15,miller16,zhou21,kawamura21,li23}.

\begin{figure}
\includegraphics[width=\linewidth]{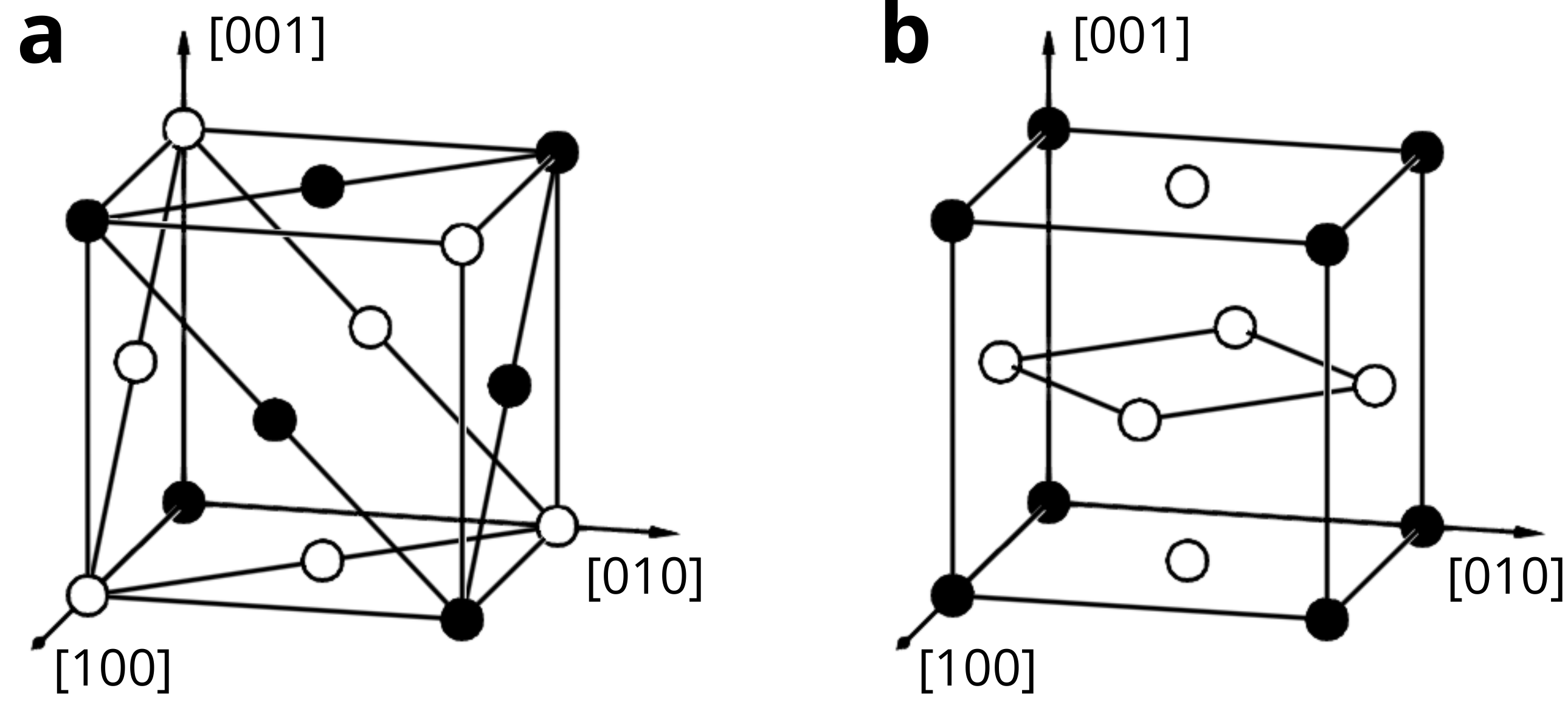}
\caption{\label{fig:structures}
    \textbf{Schematics of CuPt- and \ce{AlNi3}-type orderings.}
    \textbf{a,} The CuPt-type (L$1_1$) decoration of an fcc lattice. SRO based on this structure has been proposed for VCoNi and CrCoNi with V or Cr-rich $\bullet$ sites and complementarily depleted $\circ$ sites on alternating $(111)$ planes.
    \textbf{b,} Similarly, an \ce{AlNi3}-type (L$1_2$) unit cell.
    First-principles calculations suggest that this general form of ordering, in which V or Cr-rich $\bullet$ sites form a sublattice that minimizes nearest neighbors, should be far more energetically favorable \cite{du22}, if not the ground state \cite{schonfeld19,kostiuchenko20}, in these systems.
}
\end{figure}

Some of these reflections are consistent with the partial formation of a CuPt-type (L$1_1$) concentration wave involving the compositional enrichment and depletion of alternating $\{111\}$ (and simultaneously $\{311\}$) planes, as illustrated in Fig. \ref{fig:structures}a.
Diffuse intensities in VCoNi and CrCoNi have been interpreted to reflect modulations of V or Cr concentrations in this manner \cite{chen21,chen22,zhou22}, largely on the basis of electronic structure calculations indicating repulsive interactions between V-V and Cr-Cr neighbors, although we note that CuPt-type ordering has the same nearest-neighbor pair frequencies as a random alloy.
Some efforts have been made to support this theory with atomic-scale composition mapping \cite{chen21,zhou22,zhu23}, but, in contrast to the diffuse intensities themselves, these measurements are noisy and susceptible to local fluctuations, making it difficult to draw statistical conclusions.

\begin{figure*}
\includegraphics{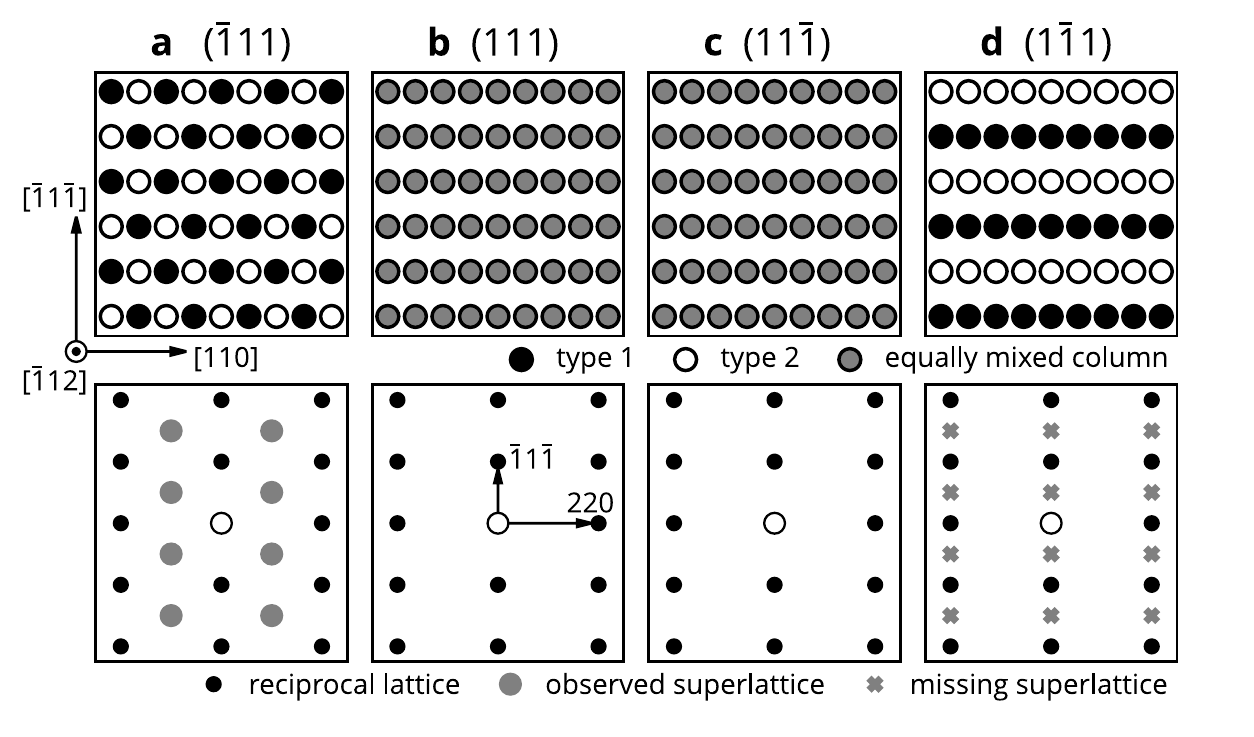}
\caption{\label{fig:cupt}
    \textbf{Diffraction of CuPt-type ordering in the $[\bar{1}12]$ ZA.}
    \textbf{a-d}, The four rotational variants of a CuPt-type structure, consisting of ordering on the denoted planes. In the top row, structures are drawn from the perspective of the $[\bar{1}12]$ ZA, with columns of atoms shaded according to the average composition. (In CuPt, types 1 and 2 would represent Cu and Pt; in recently proposed SRO, they correspond to Cr or V enrichment and depletion as in Fig. \ref{fig:structures}.)  The reciprocal-space signatures of the four variants are shown below for the same ZA. Only reflections associated with variant \textbf{a} have been reported in the discussed alloys, questioning the existence of this form of order.
}
\end{figure*}

Regardless of chemical specifics, the presence of superlattice reflections should not be regarded as incontrovertible evidence for ordering.
In fact, widespread interpretations of SRO are questionable on several accounts, such as the absence of additional expected peaks.
By the symmetry of the fcc lattice, a CuPt-type decoration can equivalently occur in four rotational variants, corresponding to order on the $(\bar{1}11)$, $(111)$, $(11\bar{1})$, or $(1\bar{1}1)$ planes, with four additional antiphase structures that are redundant for purposes of diffraction.
The four rotational variants are illustrated in the top row of Fig. \ref{fig:cupt} from the perspective of the $[\bar{1}12]$ ZA, with atomic columns shaded by composition; the reciprocal-space intensities expected from each variant are shown below for the same ZA. (Reflections were determined from the basic diffraction criterion for concentration waves \cite{khachaturyan08}, as restricted by the two-dimensional nature of TEM, and verified through simulations.)

As illustrated in Fig. \ref{fig:cupt}a, the $\frac{1}{2}\{311\}$ peaks visible in Fig. \ref{fig:diffraction} are associated only with the $(\bar{1}11)$ variant.
While the variants on the $(111)$ and $(11\bar{1})$ planes are not expected to produce additional reflections in this orientation, the $(1\bar{1}1)$-based variant depicted in Fig. \ref{fig:cupt}d should be readily visible, as it would involve composition modulation across the $(1\bar{1}1)$ planes that form the rows of atomic columns viewed in TEM.
Nonetheless, the associated $\frac{1}{2}(1\bar{1}1)$ peaks are missing from all experimental characterizations of the $[\bar{1}12]$ ZA, through either electron diffraction or the Fourier transformation of dark-field images \cite{kim15,xu15,miller16,seol20,liu21,chen21,chen22,zhou22,kayani22,zhang22,su22,seol22,li23,zhu23}.
Locations where additional reflections would be expected are circled in Fig. \ref{fig:diffraction} and marked in Fig. \ref{fig:cupt}d.
Given the quantity of material sampled across numerous studies, the absence of a variant is not statistically conceivable.

One could attempt to construct an alternative structure giving rise to only $\frac{1}{2}\{311\}$ intensities, but every $\frac{1}{2}\{311\}$ peak is related to a $\frac{1}{2}\{111\}$ spot by a $\{200\}$ reciprocal lattice vector, e.g. $\frac{1}{2}(131) - (020) = \frac{1}{2}(1\bar{1}1)$, as is the case of Fig. \ref{fig:cupt}d.
Since the diffraction criterion for concentration waves is independent of reciprocal lattice translations \cite{khachaturyan08}, any ordering that produces $\frac{1}{2}\{311\}$ peaks should also effect $\frac{1}{2}\{111\}$ intensities as long as all variants are present.

Furthermore, $\frac{1}{3}\{422\}$ and $\frac{1}{2}\{311\}$ reflections have very recently been reported in pure Cu \cite{li23}; $\frac{1}{3}\{422\}$ reflections were also previously observed in pure Ni \cite{miller16}.
Evidently, superlattice intensities in pure elements cannot represent chemical ordering and require another explanation, of which there are in fact several.

\begin{table}
\caption{\label{table:xiao94}
    \textbf{Extra reflections from fcc planar defects.}
The reciprocal-space features expected from planar defects \cite{xiao94} match recent experimental observations in concentrated alloys, as noted for each ZA.
In most cases, the intensities were originally attributed to SRO.
}
\begin{ruledtabular}
\begin{tabular}{ccc}
    ZA & Extra reflections & Observations \\
    \hline
    $[011]$ & streaking & \cite{kim15,xu15,zhang20,seol22} \\
    $[\bar{1}11]$ & $\frac{1}{3}\{422\}$ & \cite{kim15,xu15,miller16,kawamura21,zhou21,chen22,li23,kayani22} \\
    $[\bar{1}12]$ & $\frac{1}{2}\{311\}$ & \cite{kim15,xu15,miller16,seol20,liu21,chen21,chen22,zhou22,kayani22,zhang22,su22,seol22,li23,zhu23} \\
    $[013]$ &  $\frac{1}{2}\{311\}$ & \cite{miller16,kayani22} \\
\end{tabular}
\end{ruledtabular}
\end{table}

One is the presence of nanoscale planar defects.
Forbidden reflections expected from stacking faults or nanotwins in an fcc lattice \cite{xiao94} are listed in Table \ref{table:xiao94}.
Remarkably, these are the exact features that have been reported in the $[011]$, $[\bar{1}12]$, $[\bar{1}11]$, and $[013]$ ZAs of concentrated alloys, offering an alternative explanation for the experimental findings described above.
As structural defects break the symmetry of the reciprocal lattice, it is possible for $\frac{1}{2}\{311\}$ peaks to appear without $\frac{1}{2}\{111\}$ counterparts.
Of course, the obvious objection to this hypothesis is that most imaged samples appeared to contain no such imperfections in the examined regions.
Considering that stacking faults and related structures are usually quite visible under TEM, some explanation of how presumably nanoscale planar defects could otherwise escape detection would be required to prove their generation of the discussed features.

While clearly visible stacking faults produce sharp superlattice reflections \cite{xiao94}, the diffuseness of the discussed intensities could imply the presence of smaller, less readily detected defects, just as SRO causes faint reflections compared to the sharp superlattice peaks resulting from long-range order (LRO).
Both faulted (Frank) loops and stacking-fault tetrahedra should produce the extra reflections listed in Table \ref{table:xiao94} and, if small enough, could plausibly escape direct recognition.
In many recent studies, superlattice intensities have been associated with tiny localized features, which could correspond to defect structures.
Moreover, two recent investigations have connected the enhancement of diffuse intensities to mechanical deformation \cite{seol22} and irradiation \cite{su22}, while brighter reflections have been found near a crack tip \cite{kim15}.
Both deformation and irradiation are well known to induce planar defects at the expense of chemical order, supporting this hypothesis---in fact, the intensity of some peaks has been directly correlated the density of planar defects \cite{seol22}.

Most observations in the $[011]$ ZA support the absence of detectable SRO---CuPt-type ordering should produce additional superlattice reflections in this ZA, which are generally not observed \cite{zhang20,chen21,zhou22}.
(Faint intensities at $\frac{1}{2}\{111\}$ positions were suggested by Ref. \cite{chen22}, but the signal in this region seems comparable to the background noise level.
Clearer $\frac{1}{2}\{111\}$ intensities in the $[011]$ ZA have been proposed following the application of a novel post processing algorithm \cite{hsiao22}, although we believe this technique may require further discussion of a technical nature beyond the scope of this comment.)
Several studies have instead reported streaking in this ZA \cite{kim15,xu15,zhang20,seol22}, which would also be consistent with planar defects \cite{xiao94}. Others have found no extra features, but the absence of observation in specific instances would hardly be surprising given variation in sample preparation and the inherently local nature of the proposed defects.

Another theory is that the extra reflections are merely artifacts caused by ``relrod spiking" from higher-order Laue zone (HOLZ) diffraction \cite{li23}, which could in principle account for the locations of most reported peaks, although the HOLZ intensities predicted by kinematical theory are negligible \cite{miller16}.
While dynamical scattering could theoretically contribute to forbidden reflections, it is not immediately clear how this would occur in the sample geometries used in the literature, which were too thin to even produce Kikuchi diffraction.
Streaking in the $[011]$ ZA \cite{kim15,xu15,zhang20,seol22} and the specific peaks observed in the $[013]$ ZA \cite{miller16,kayani22} may also be more consistent with faulting \cite{xiao94}.

It has been additionally suggested that surface steps could produce the extra reflections \cite{miller16,li23}, as has been demonstrated for $\frac{1}{3}\{422\}$ intensities in the $[\bar{1}11]$ ZA \cite{cherns74}.
However, it is less clear if this mechanism is consistent with observations in other ZAs.
Additionally, surface-step reflections have been primarily observed in deposited thin films with clear step contours, which are not apparent in recent observations of differently processed samples.

Somewhat before recent interest, $\frac{1}{2}\{311\}$ reflections in the $[\bar{1}12]$ ZA of an Al$_{0.5}$CrFeCoNiCu alloy were suggested to possibly originate from thermal diffuse scattering \cite{xu15}. However, it has also been shown that at least $\frac{1}{3}\{422\}$ reflections remain essentially unchanged at liquid-nitrogen temperatures \cite{miller16}.
Alternatively, scattering from static lattice displacements has been proposed as another potential source of extra intensities \cite{zhou21,kawamura21}.
We note that this phenomenon can break the symmetry of the reciprocal lattice \cite{cook69b}, although a specific mechanism by which it could produce the discussed observations has not been established.
Like SRO, static displacement scattering also could not explain the intensities observed in pure elements.

While the connection between the extra reflections and SRO is questionable on a purely experimental basis, we further note that the proposed structures are largely inconsistent with the bonding principles predicted by standard density-functional theory, either directly or through parameterized interatomic potentials.
As previously noted \cite{zhou22,du22}, CuPt-type ordering is clearly energetically unfavorable in otherwise similar V-Ni and Cr-Ni alloys.
In these systems, experimental SRO has been primarily interpreted in terms of \ce{AlNi3} (L$1_2$, see Fig. \ref{fig:structures}) or \ce{Al3Ti}-type (DO$_{22}$) concentration waves \cite{schonfeld88,caudron92,schonfeld94,lebolloch00}, both of which minimize nearest neighbors among the ordering solute.

First-principles calculations consistently indicate that VCoNi and CrCoNi should order similarly to the aforementioned binaries, with VCoNi clearly favoring an \ce{AlNi3}-type V sublattice \cite{kostiuchenko20} and similar, though not identical, preferences noted for CrCoNi \cite{walsh21,du22,ghosh22}.
While SRO may differ from the LRO ground state \cite{khachaturyan08}, the underlying interactions are expected to be comparable and there is essentially no indication of any energetic driving force for the formation of CuPt-type ordering.

In practice, the basic predictions of electronic structure calculations are largely supported by diffuse X-ray scattering in CrFeCoNi, which reveals an incipient \ce{AlNi3}-type Cr sublattice \cite{schonfeld19} after long-term aging below the order-disorder transition temperature.
Moreover, VCoNi alloys readily form fully ordered \ce{AlNi3}-type domains (see Fig. \ref{fig:structures}), which were observed alongside the nominally disordered regions characterized by Ref. \cite{chen21}.
It would be unexpected for this material to host SRO corresponding to an unrelated structure immediately adjacent to the theoretically predicted LRO.

A few studies have nonetheless tried to reconcile experimental observations with theoretical predictions.
In particular, local instances of ordering on $\frac{1}{2}\{311\}$ planes were identified in high-temperature thermodynamic simulations of CrCoNi parameterized by a carefully developed ``neural network" interatomic potential \cite{du22}.
However, the CuPt-type structure was noted to be energetically unfavorable and it is unclear if these regions represent anything beyond random fluctuations.
Their equilibrium frequency does not vary with temperature above the order-disorder transition and equivalent instances of theoretically favorable $\{100\}$ and $\{110\}$-based motifs are consistently more prevalent, even though their associated peaks are not found experimentally.

Clearer agreement has been found between experimental diffraction patterns and simulations \cite{hsiao22} for certain previously generated theoretical structures \cite{walsh21}.
These, however, were selected from a large collection of tiny configurations that were created to statistically explore highly speculative ordering principles rather than represent realistic chemical environments.
Consequently, individual structures probably contained configurational fluctuations that, given the small cell size, could lead to various extra reflections.

Altogether, there seems to be little theoretical basis for any form of SRO consistent with the electron diffraction of VCoNi, CrCoNi, and other similar alloys, while reported features consistently match those expected from symmetry-breaking effects such as changes in the stacking sequence.
On the weight of the circumstantial evidence, we believe that assumptions of ordering should be revisited and the possibility that other phenomena produced the observed diffuse intensities deserves further investigation.
This is not to say that any experimental sample necessarily lacked SRO, simply that it may not be definitively detected by the employed techniques.

\begin{acknowledgments}
This work was supported by the US Department of Energy, Office of Science, Office of Basic Energy Sciences, Materials Sciences and Engineering Division under contract No. DE-AC02-05CH11231 as part of the Damage-Tolerance in Structural Materials (KC13) program.
Work at the Molecular Foundry was supported by the US Department of Energy, Office of Science, Office of Basic Energy Sciences under the same contract.
Resources provided by award No. BES-ERCAP0021088 of the National Energy Research Scientific Computing Center, a US Department of Energy Office of Science User Facility operated under the same contract, were also used.
FW additionally thanks Q. Yu for insightful conversations.
\end{acknowledgments}

\section*{Competing interests}
The authors declare no competing interests.

\end{document}